\documentclass[%
 reprint,
superscriptaddress,
 amsmath,amssymb,
 aps,
prb,
floatfix
]{revtex4-2}

\usepackage{graphicx}
\usepackage{dcolumn}
\usepackage{natbib}
\usepackage{comment}
\usepackage{braket}
\usepackage{amsmath}
\usepackage{multirow}
\usepackage{array}
\newcolumntype{C}[1]{>{\centering\let\newline\\\arraybackslash\hspace{0pt}}m{#1}}
\usepackage{booktabs}
\usepackage{color,soul}
\usepackage{epstopdf}

\begin{document}
\preprint{APS/123-QED}

\title{Exciton-exciton Interactions -- A Quantitative Comparison Between Complimentary Phenomenological Models}
\author{Pradeep Kumar}
\affiliation{Department of Physics, Indian Institute of Science Education and Research Bhopal, Bhopal 462066, India}
\author{Bhaskar De}%
\affiliation{Department of Physics, Indian Institute of Science Education and Research Bhopal, Bhopal 462066, India}
\author{Rishabh Tripathi}
\affiliation{Department of Physics, Indian Institute of Science Education and Research Bhopal, Bhopal 462066, India}
\author{Rohan Singh}
\email{rohan@iiserb.ac.in}
\affiliation{Department of Physics, Indian Institute of Science Education and Research Bhopal, Bhopal 462066, India}
\affiliation{JILA, University of Colorado \& National Institute of Standards and Technology, Boulder, Colorado 80309-0440, USA}
\affiliation{Department of Physics, University of Colorado, Boulder, Colorado 80309-0390, USA}

\date{\today}

\begin{abstract}
Many-body interactions such as exciton-exciton interactions significantly affect the optical response of semiconductor nanostructures.
These interactions can be rigorously modeled through microscopic calculations.
However, these calculations can be computationally intensive and often lack physical insights.
An alternative is to use phenomenological many-body-interaction models such as the modified optical Bloch equations and the anharmonic oscillator model.
While both these models have separately been used to interpret experimental data, to the best of our knowledge, an explicit and direct correspondence between these models has not been established.
Here, we show the empirical equivalence between these two complimentary models through two-dimensional coherent spectroscopy simulations.
A quantitative correspondence between the parameters used to incorporate the exciton-exciton interactions in these two models are obtained.
We also perform a quantitative comparison of these phenomenological models with experiments, which highlights their usefulness in interpreting experimental results.
\end{abstract}


\maketitle



\section{\label{sec:intro}Introduction}
The bound state of an electron-hole pair, known as exciton, dominates the near-bandgap optical response of semiconductor nanostructures.
The Coulomb interactions that bind the electron-hole pair also influence other excitons, resulting in strong many-body interactions (MBIs).
These MBIs in quantum wells have previously been demonstrated by excitation-density-dependent optical measurements \cite{chemla2001many, wachter2002coherent}.
More recently, similar phenomena have been observed in other  2D nanomaterials, including perovskite nanoplatelets \cite{wu2015excitonic, thouin2019enhanced}, transition metal dichalcogenides (TMDCs) \cite{jakubczyk2019coherence, boule2020coherent, katsch2020exciton, erkensten2021exciton}, and Rydberg excitons in solids \cite{kazimierczuk2014giant, semkat2021quantum}.
MBIs play a key role in explaining the origin of electronic coherent coupling in quantum-dot states \cite{lenihan2002raman, martin2018inducing}, quantum wells \cite{ferrio1998raman,kasprzak2011coherent,nardin2014coherent, moody2014coherent}, and TMDCs \cite{singh2014coherent, purz2021coherent}. 
While these novel materials are of interest for quantum information processing, manipulating and controlling these MBIs will be critical for scalable applications.
Furthermore, a significant enhancement of photocatalytic efficiency was attributed to MBIs recently \cite{mukherjee2023many}.
Thus, quantifying these MBIs can help to optimize the performance of optoelectronic devices.

The effects of the MBIs are manifested in the coherence decay dynamics of excitons.
Typically, nonlinear techniques like four-wave mixing (FWM) \cite{Shah1999} and two-dimensional coherent spectroscopy(2DCS) \cite{turner2011invited, tollerud2017coherent, li2023optical} are used to study the coherent dynamics of these nanomaterials' optical response \cite{li2006many, webber2015role, moody2015intrinsic}.
Experimental observation of these many-body signatures can be interpreted through various formalisms such as local-field correction (LFC) \cite{leo1990effects, wegener1990line, kim1992unusually}, excitation-induced dephasing (EID) \cite{honold1989collision, wang1993transient, hu1994excitation}, excitation-induced shift (EIS) \cite{mitsunaga1992excitation, shacklette2002role}, and biexciton formation \cite{mayer1994evidence, stone2009two}.

The most rigorous method to incorporate these MBIs is through a comprehensive many-body treatment, referred to as the microscopic model \cite{axt1994dynamics}.
Although based on a sound theoretical foundation, the microscopic model can be computationally challenging to execute and often provides limited physical insight into the underlying phenomena \cite{bolton2000demonstration}.
To overcome these limitations, phenomenological models were developed as simpler and physically-intuitive alternatives.

One of the phenomenological few-level models is the modified optical Bloch equations (MOBEs) \cite{shacklette2003nonperturbative}.
The excitons are primarily treated as a two-level system (fermions) and the OBEs are used to describe their interaction with light.
The MOBEs are obtained by incorporating the MBIs as three excitation-dependent modifications to the OBEs.
These additional terms include two terms correspond to change in the resonance energy (EIS) and dephasing rate or homogeneous linewidth (EID) with the excitation density.
The third additional term is the LFC that incorporates changes in the effective optical electric field due to macroscopic polarization.
The EIS and EID effects were measured as a change in the exciton resonance energy \cite{peyghambarian1984blue, chemla1989excitonic, manzke1998density} and dephasing rate \cite{honold1989collision, wang1993transient}, respectively, with an increase in the excitation density whereas the LFC was primarily invoked to explain the origin of the so-called negative-delay signal in two-pulse FWM experiments \cite{leo1990effects, wegener1990line, kim1992unusually}.
The MOBEs provide a good qualitative explanation for the experimental signatures arising from MBIs.
However, MOBEs need to be solved numerically \cite{shacklette2002role}, which inhibits their use for quantitative analysis of data.

An alternative phenomenological framework to understand the nonlinear optical response of the excitons in the presence of MBIs is the anharmonic oscillator (AO) model \cite{kuwata1997parametric, svirko1999four, inoue2000renormalized, rudin2001anharmonic}.
Here, MBIs between excitons, which are composite bosons, are introduced in the form of anharmonic corrections to the harmonic potential of non-interacting excitons.
While an ideal bosonic system is perfectly linear and cannot be saturated, the anharmonicity results in a nonlinear response of the excitons.
Changes in the energy spacing, dephasing rate and transition dipole moment of higher excited states account for EIS, EID, and phase-space filling (PSF), respectively, in this model.
A significant advantage of the AO model is that we can analytically calculate the nonlinear optical response by solving the relevant OBEs perturbatively.
Quantification of the MBIs by fitting the experimental data with the AO model has been demonstrated previously \cite{singh2016polarization}.

Both the aforementioned phenomenological models have been separately used to explain and interpret the effects of MBIs in experiments \cite{kuwata1997parametric, li2006many, singh2016polarization}.
While these models are based on microscopic descriptions and can be understood through fermionic (MOBE) \cite{victor1995microscopic} and bosonic (AO) \cite{svirko1999four} frameworks of interacting excitons, one would expect the convergence of these two in the limit of a large ensemble and at low excitation density.
However, to the best of our knowledge, a direct correspondence between these two complimentary models has not been established. 
For instance, it is not known if the parameters used to quantify MBIs in the two models are equivalent.
Thus, a quantitative comparison of the two models is essential for gaining a comprehensive understanding of the MBIs and their effect on the nonlinear-optical response.
Specifically, these findings will enable us to compare the relative strength of MBI in different nanomaterials. 

In this work, we  have established an empirical equivalence between these two complimentary MBI models through 2DCS simulations.
In Sec. \ref{sec:mobe} we have simulated 2D spectra for a range of excitation densities using the MOBEs.
Slices from the simulated spectra are fit to those obtained from calculations for the AO model in Sec. \ref{sec:ao}.
We find an excellent match between the peak lineshape in 2D spectra from both approaches for a range of excitation densities and MBI strengths, which demonstrates their equivalence.
Furthermore, we discuss the correspondence between the MBI parameters used in the two approaches.
In Sec. \ref{sec:expt}, we use both the MBI models to quantitatively reproduce the excitation-density dependence of lineshapes in 2DCS experiments which were performed on GaAs QWs.
We provide a summary and outlook of our work in Sec. \ref{sec:end}.

\section{\label{sec:mobe}Modified-Optical-Bloch-Equations simulation}

\begin{figure}[t]
\includegraphics{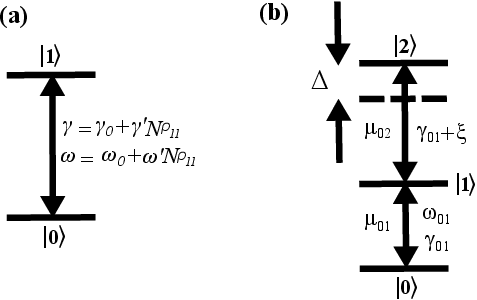}
\caption{(a)Exciton as a two-level system used for the MOBE calculations with excitation-density-dependent dephasing rate $\gamma$ and resonance frequency $\omega$.
Here $\ket{0}$ and  $\ket{1}$ represent the ground and excited states, respectively.
(b) Energy level scheme for anharmonic exciton ladder is illustrated.
$\ket{0}$ : ground state, $\ket{1}$ : single-exciton state, $\ket{2}$: two-exciton state, $\gamma_{01}$: dephasing rate of the $\ket{0}$$\leftrightarrow$$\ket{1}$ transition, $\xi$: excitation induced dephasing (EID), $\Delta$: excitation induced shift (EIS)}
 \label{fig:fig1_MOBE _AO_model}
\end{figure}

Figure \ref{fig:fig1_MOBE _AO_model}(a) shows the exciton as a two-level system.
The ground and excited states are denoted as $\ket{0}$ and $\ket{1}$, respectively.
The MOBEs incorporate EID and EIS through a linear dependence of dephasing rate $\gamma$ and resonance frequency $\omega$ with excitation density
\begin{subequations}
    \begin{eqnarray}
        {\gamma} & = &{\gamma_{0}}+ \gamma'N{\rho_{11}},\label{a}
        \\
        {\omega} & = &{\omega_{0}}+ \omega'N{\rho_{11}}. 
        \label{b}
    \end{eqnarray}
\end{subequations}
Here, the dephasing rate $\gamma$ and the resonance frequency $\omega$ comprise the excitation-independent terms ($\gamma_{0}$, $\omega_{0}$) and the excitation dependent terms ($\gamma'N\rho_{11}$, $\omega'N\rho_{11}$).
$N\rho_{11}$ is the excited state population density with density of states $N$ and the density matrix element $\rho_{11}$ corresponding to the occupation probability of $\ket{1}$.
The parameters $\gamma'$ and $\omega'$ quantify the strength of EID and EIS, respectively.  
The MOBEs are obtained from the OBEs after incorporating the above mentioned excitation-density-dependent terms:
\begin{subequations}
    \begin{align}
       \dot{\rho}_{11} & = {-\Gamma_1}\rho_{11} + \frac{i}{\hbar}(\mu E)(\rho_{01}- \rho_{10})\\
       \dot{\rho}_{01} & =-(\gamma_{0}+\gamma'N\rho_{11})\rho_{01}+i(\omega_{0} +\omega'N\rho_{11})\rho_{01}\\
      			&\quad +\frac{i}{\hbar}(\mu E)(\rho_{11}-\rho_{00}), \nonumber
    \end{align}
\end{subequations}      
where $\rho_{jk}$ are the density matrix elements, in which $j = k$ and  $j\neq k$  correspond to  population and coherence terms, respectively, $\Gamma_{1}$ is the population decay rate of the exciton and $\mu$ is the transition dipole moment of $\ket{0}\leftrightarrow\ket{1}$ transition.
The electric field of the excitation light is
\begin{equation}
\label{eq:electric_field}
    E(\mathbf{r},t) = \frac{\mathcal{E}(t)}{2}[e^{i(\omega_{L} t-\mathbf{k}\cdot\mathbf{r})} + c.c.],
\end{equation}
where $\mathcal{E}(t)$ is time-dependent electric-field amplitude, $\omega_{L}$ is the frequency of the excitation light, $\mathbf{k}$ is the wavevector and $\mathbf{r}$ is position.
Here, we have not included the LFC term since its primary effect is to produce the negative time delay transient four-wave mixing signal (TFWM), which is analogous to two quantum coherence signals in 2DCS; this signal is not relevant for the current work.

\begin{figure}[t]
\includegraphics{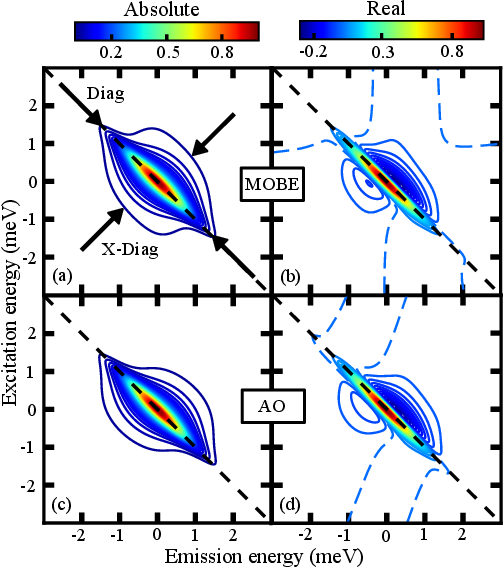}
\caption{The (a) absolute and (b) real part of the simulated 2D spectra using MOBEs for a pulse area of $3\times10^{-3} \pi$, $N\gamma'$ = 100 meV and $N\omega'$ = 50 meV. 
The (c) absolute and (d) real part of the simulated 2D spectra using the best-fit parameters for the AO model.
Dashed diagonal line indicates equal excitation and emission energy amplitudes. 
The cross-diagonal (X-Diag) and Diagonal (Diag) directions are indicated by arrows in (a).
The dashed contours in (b) and (d) show the zero level.}
\label{fig:fig2_2Dspectra}
\end{figure}

\begin{figure}[t]
\includegraphics{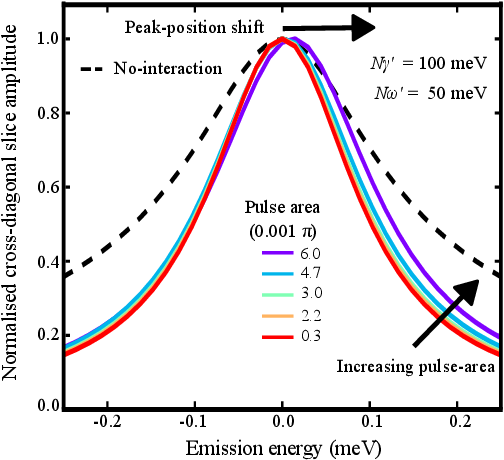}
\caption{Pulse area dependent cross-diagonal slices of absolute value 2D spectra obtained from MOBE model are shown.
Excitation induced dephasing (EID) causes a broadening of the cross-diagonal slices with increasing pulse area. 
A slight blue shift in the peak of the cross-diagonal slices and asymmetric line shape indicates excitation induced shift (EIS). 
The black arrows indicate increasing pulse area.
The dashed line shows the cross-diagonal slice in the absence of MBIs.}
\label{fig:fig3_crossdiagonal_final}
\end{figure}

We have solved the MOBEs numerically to obtain the two-pulse TFWM signal \cite{leitenstorfer1994excitonic, shacklette2003nonperturbative}.
A pair of excitation pulses with wavevectors $\bf{k_1}$ and $\bf{k_2}$ are considered.
The time delay between the pulses is $\tau$ with the $\bf{k_1}$ arriving first.
The FWM signal is emitted during time $t$ after the arrival of the second pulse.
The homogeneous TFWM signal $S_{\rm{hom}}(\omega,\tau, t)$ for resonance frequency $\omega$ is obtained by extracting the third-order coherence term $\rho^{(3)}_{01}(\tau, t)$ in the phase-matching direction  $2 {\bf k_{2}} - {\bf k_{1}}$.
The total TFWM signal for the inhomogeneous distribution is obtained by taking the sum of the signal from all the frequency components, $S_{\rm{inhom}}(\tau, t)=\sum_{i} p_{i} S_{\rm{hom}}(\omega_i,\tau, t)$, where $p_i$ is the normalized probability of having an oscillator with resonance frequency $\omega_i$.
We consider a Gaussian distribution with standard deviation $\sigma$ for the inhomogeneity.
The MOBEs are solved for transform-limited excitation pulses with a Gaussian envelope and 100 fs full-width half-maximum (FWHM).
The intensities of both the excitation pulses are kept equal to each other.
We quantify the pulse intensity in terms of square of the pulse area $\Theta = \dfrac{1}{\hbar}\int_{-\infty}^{\infty} \mu \mathcal{E}(t) dt$ \cite{allen1987optical}.
Simulated 2D spectra $\mathcal{S}_{\rm{inhom}}(\omega_{\tau}, \omega_{t})$ are obtained by taking a Fourier transform of the time-domain FWM signal $S_{\rm{inhom}}(\tau, t)$ along the axes $\tau$ and $t$.
Figures \ref{fig:fig2_2Dspectra}(a) and \ref{fig:fig2_2Dspectra}(b) show the absolute and real part, respectively, of the 2D spectra obtained for $\Theta = 3\times10^{-3} \pi$.
Assuming a non-interacting two-level system, the widths of the peak along the diagonal and cross-diagonal directions indicate inhomogeneous ($\sigma$)  and homogeneous ($\gamma$) linewidths, respectively \cite{siemens2010resonance}. 
The parameters corresponding to inhomogeneous ($\sigma$) and excitation-independent homogeneous linewidth ($\gamma_{0}$) used in the simulations are 0.5 meV and 0.1 meV, respectively.
We have taken $N\gamma' = 100$ meV and $N\omega' = 50$ meV.
The dispersive lineshape of the real part of 2D spectrum in Fig. \ref{fig:fig2_2Dspectra}(b) is attributed to EIS \cite{li2006many}.

We performed MOBEs-based simulations for a range of pulse areas within the $\chi^{(3)}$ regime $\left(\int |S_{\rm{inhom}}(\tau,t)|^2 \,dt \propto \Theta^6\right)$ to measure the excitation-density-dependence of 2D spectra. 
An increase in the width of the cross-diagonal slices of the absolute 2D spectra, as shown in Fig. \ref{fig:fig3_crossdiagonal_final}, highlights the effect of EID \cite{honold1989collision, moody2015intrinsic}.
In addition to the broadening, the cross-diagonal slices show a slight blue shift and asymmetric lineshape at higher pulse areas, which can be attributed to EIS \cite{peyghambarian1984blue,manzke1998density}.
Figure \ref{fig:fig3_crossdiagonal_final} also shows that  FWM signal decay rate in the presence of MBI is slower than in the case of no interaction; this apparent anomaly will be discussed later.
While it is straightforward to qualitatively reproduce the experimental effects of EID and EIS through the MOBEs, it is difficult to quantify this effect since these numerical calculation take significant computational time.

\begin{figure*}[t]
\includegraphics{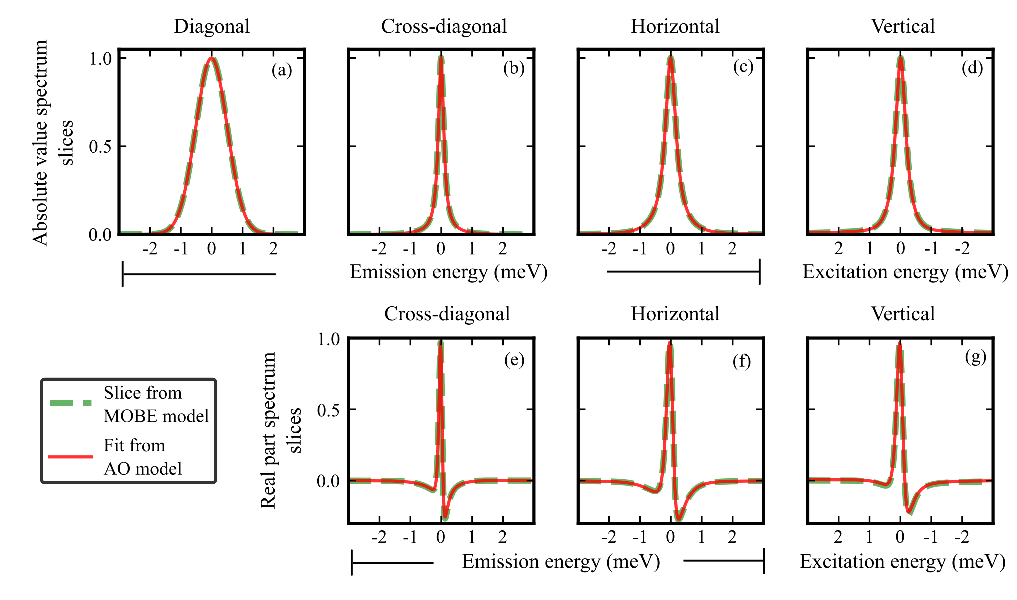}
\caption {Slices obtained from MOBE simulations (green dashed line) and the corresponding fits using the AO model (solid red line) are shown.
Slices for the MOBE simulation are taken from the absolute value (Fig. \ref{fig:fig2_2Dspectra}(a)) and real part (Fig. \ref{fig:fig2_2Dspectra}(b)) 2D spectra, respectively.
The particular slice is indicated by the label above each plot.}
\label{fig:fig4_slices}
\end{figure*}

\section{\label{sec:ao}Anharmonic model fits}

An alternative approach is to treat excitons as interacting bosons.
Interaction terms are added to the perfectly harmonic oscillator hamiltonian, which results in the anharmonic ladder of states \cite{svirko1999four} as shown in Fig. \ref{fig:fig1_MOBE _AO_model}(b) with ground $\ket{0}$, one-exciton $\ket{1}$, and two-exciton $\ket{2}$ states. 
$\mu_{01}$ and $\mu_{02} = \sqrt{2}\mu_{01}$ are the transition dipole moments for transitions $\ket{0}$$\leftrightarrow$$\ket{1}$ and $\ket{1}$$\leftrightarrow$$\ket{2}$, respectively.
The anharmonic ladder is truncated to the two-exciton state in Fig. \ref{fig:fig1_MOBE _AO_model}(b) since only these states contribute to the FWM signal in the $\chi^{(3)}$ regime. 
The anharmonic terms introduce an asymmetry between the $\ket{0}$$\leftrightarrow$$\ket{1}$ and $\ket{1}$$\leftrightarrow$$\ket{2}$ transitions.
This asymmetry can be quantified by the phenomenological terms $\Delta$ and $\xi$, which are similar to the EIS and EID effects discussed above and produces a nonzero optical nonlinear FWM signal.
We have ignored the phase-space-filling effect here, since it was previously shown to be insignificant \cite{singh2016polarization}.
The analytical solution of the FWM signal in the perturbative regime, considering $\delta$-function excitation pulses, is 
\begin{eqnarray}
\label{eq:3_SAO}
S_{AO}(\tau, t)=A e^{-i\omega_{01}(\tau-t)-\gamma_{01}(\tau+t)-\frac{\sigma^2}{2}{(\tau-t)^2}} \nonumber \\ \left[1-e^{(i\Delta-\xi)t}\right],
\end{eqnarray}
where A is the FWM-signal amplitude, $\omega_{01}$, $\gamma_{01}$, and $\sigma$ are the resonance frequency, dephasing rate or homogeneous line width, and inhomogeneous line width, respectively, of the $\ket{0}$$\leftrightarrow$$\ket{1}$ transition.
We have considered equal inhomogeneity for both $\ket{0}$$\leftrightarrow$$\ket{1}$ and $\ket{1}$$\leftrightarrow$$\ket{2}$ transitions.
This analytical calculation can be used to quantify these effects through a nonlinear fitting procedure \cite{singh2016polarization}.

We explore the correspondence between the two models by fitting multiple slices of the absolute and real part of 2D spectrum obtained from MOBE simulations with those obtained from the AO model.
The fitting procedure is repeated for a series of excitation densities and various values of $N\gamma'$ and $N\omega'$.
As an example, Fig. \ref{fig:fig4_slices} shows the slices from the 2D spectra in Figs. \ref{fig:fig2_2Dspectra}(a) and \ref{fig:fig2_2Dspectra}(b) (dashed line) along with the fits obtained from the AO model (solid line).
We find an excellent match between the simulated and fit slices.
There are slight variations in the obtained values of the fit parameters depending on the energy range chosen for the fitting procedure.
In order to avoid this bias and estimate the uncertainty in the fit parameters, we repeat the fitting procedure for several energy ranges from 10 to 20 meV.
Subsequently we will report the mean and standard deviation of the parameters obtained for these different energy ranges as the best-fit parameters when fitting MOBE simulation results with the AO model.
We obtain the following best-fit parameters for the current case -- $\gamma_{01} = 102.70 \pm 0.05$ $\mu$eV, $\sigma = 499.7 \pm$ 0.2 $\mu$eV, $\Delta = 1.82 \pm 0.05$ $\mu$eV, $\xi = 3.73 \pm 0.09$ $\mu$eV.
Simulated absolute and real-part spectra of AO model using these best-fit parameters are shown in Figs. \ref{fig:fig2_2Dspectra}(c) and \ref{fig:fig2_2Dspectra}(d), respectively.

\begin{figure}[t]
\includegraphics{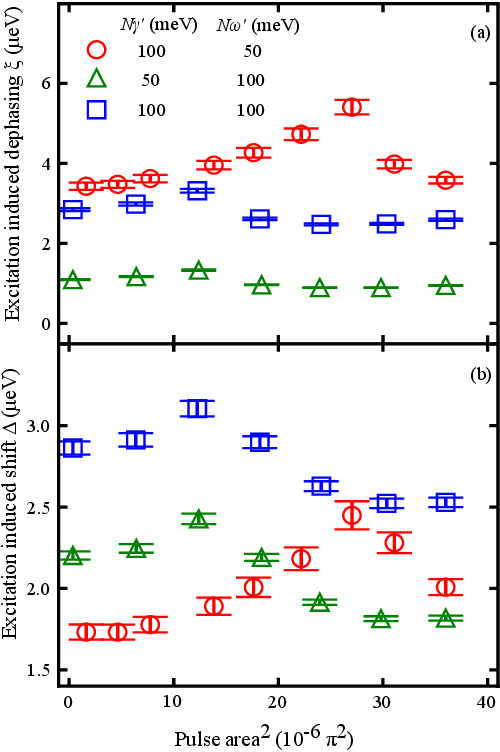}
\caption{The symmetry-breaking fit parameters in the AO model -- EID ($\xi$) in (a) and EIS ($\Delta$) in (b) measured as a function of excitation densities for different values of $N\gamma'$ and $N\omega'$.
The error bars are standard deviation in parameters vaule obtained by fitting 2D spectra in multiple energy ranges.}
\label{fig:fig5_EIDEIS}
\end{figure}

The symmetry-breaking fit parameters $\xi$ and  $\Delta$ of the AO model are shown in Figs. \ref{fig:fig5_EIDEIS}(a) and \ref{fig:fig5_EIDEIS}(b), respectively.
These parameters don't exhibit any clear dependence on the excitation density; this behaviour is observed for other combinations of $N\gamma'$ and $N\omega'$ too.
Because of the random variation with excitation density, we quantify the values of $\Delta$ and $\xi$ by taking a weighted average of the results for all excitation densities; these are summarized in Table \ref{tab:symm}.
These values change with the values of $N\gamma'$ and $N\omega'$ used in the MOBE simulation.
We find that both $\Delta$ and $\xi$ vary when either $N\gamma'$ or $N\omega'$ is changed.
While these variations seem random, we note that the ratio $\dfrac{\xi}{\Delta}$ matches perfectly with the value of $\dfrac{\gamma'}{\omega'}$.
We thus conclude that although the parameters $\xi$ and $\Delta$ are not perfectly separable, the fitting procedure gives a good estimate of the relative strength of EID and EIS.

\begin{table}
\caption{\label{tab:symm}Symmetry breaking fit parameters in the AO model.}
\begin{center}
\begin{tabular}{C{3em} C{3em} C{6em} C{6em} C{6em}}
\toprule
\textbf{$N\boldsymbol{\gamma'}$}&\textbf{$N\boldsymbol{\omega'}$}&
\textbf{$\boldsymbol{\xi}$}&\textbf{$\boldsymbol{\Delta}$}&
      \multirow{2}{1em}{$\dfrac{\xi}{\Delta}$}\\
      (meV) & (meV) & ($\mu$eV) & ($\mu$eV) & \\
\midrule
100 & 100  & 2.65 $\pm$ 0.01 & 2.70 $\pm$ 0.01 & 0.981 $\pm$ 0.005 \\
50 & 100  & 1.000 $\pm$ 0.003 & 1.978 $\pm$ 0.007 & 0.506 $\pm$ 0.002 \\
100 & 50  & 3.72 $\pm$ 0.03 & 1.86 $\pm$ 0.01 & 2.00 $\pm$ 0.02\\
 \bottomrule
\end{tabular}
\end{center}
\end{table}

\begin{figure}[t!]
\includegraphics{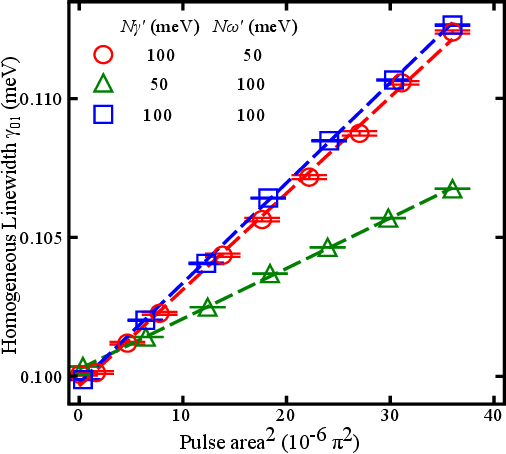}
\caption{Linear variation of homogeneous linewidth with  excitation density for different combinations of $N\gamma'$ and  $N\omega'$. 
Linear fits to data are shown by the dashed lines.}
\label{fig:fig6_Homogeneous line width_AO}
\end{figure}

The excitation-independent parameters $\xi$ and $\Delta$ do not affect the overall width and position of the peak and, consequently, cannot model the broadening and shift of the cross-diagonal slices shown in Fig. \ref{fig:fig3_crossdiagonal_final}.
This goal is achieved by introducing parameters $\gamma_{01}(N_x)$ and $\omega_{01}(N_x)$ that depend on the excitation density $N_x$.
Figure \ref{fig:fig6_Homogeneous line width_AO} shows an increase in the homogeneous linewidth $\gamma_{01}$ with the excitation density.
We quantify the excitation-density dependence using the relation
\begin{equation}
\label{eq:4_gamma01}
\gamma_{01}(N_{x})= \gamma_{01}{(0)} + \gamma_{EID} N_{x},
\end{equation}
where  $\gamma_{01}{(0)}$ is the zero-excitation-density homogeneous linewidth of the $\ket{0}$$\leftrightarrow$$\ket{1}$ transition, and $\gamma_{EID}$ is the rate of change of homogeneous line width with the excitation density $N_{x}$ \cite{honold1989collision,singh2013anisotropic}.
The fit results are summarized in Table \ref{tab:homo}.
We find that the slope $\gamma_{EID}$ is directly proportional to $N\gamma'$ and does not vary significantly with the value of $N\omega'$.
Furthermore, the offset $\gamma_{01}(0)$ is independent of the MBI parameters used in the MOBEs and equals the excitation-independent term $\gamma_0$.

\begin{table}[h]
\caption{\label{tab:homo}Excitation density dependence homogeneous line width ($\gamma_{01}$).}
\begin{center}
\begin{tabular}{cccc}
\toprule
      \textbf{$N\boldsymbol{\gamma'}$} &\textbf{$N\boldsymbol{\omega'}$}& \textbf{$\boldsymbol{\gamma_{EID}}$}&\textbf{$\boldsymbol{\gamma_{01}(0)}$}\\ 
      (meV) & (meV) & {$\left(10^6 \pi^{-2} \,\textrm{meV}\right)$} & ($\mu$eV)\\
      \midrule
      100 & 100 & 360  $\pm$ 2 & 99.76 $\pm$ 0.04 \\
      50 & 100 & 181.0  $\pm$ 1.3  & 100.26 $\pm$ 0.03 \\
      100 & 50 & 346  $\pm$ 5 & 99.65  $\pm$ 0.08\\
      \bottomrule
    \end{tabular}
 \end{center}
\end{table}

Now let's shift our focus to an intriguing feature shown in Fig. \ref{fig:fig3_crossdiagonal_final}.
The cross-diagonal slice is narrower when the MBIs are included than without them. 
This narrowing indicates that the decay rate of the FWM signal is slower in presence of EID than in its absence.
It is worth noting that a previous observation also supports this behavior of narrowing homogeneous linewidth when MBIs are present, specifically in the context of the bosonic framework of excitons \cite{singh2016polarization}.
In that case, the interference between two different quantum mechanical pathways results in a narrower linewidth.
Since the MOBEs are solved through numerical methods, we cannot provide an intuitive explanation for this narrowing in this particular case.
However, the similar behaviour further demonstrates the equivalence of incorporating the MBIs either through the MOBEs or the AO model.
Additionally, we want to emphasize that $\gamma_0$, is not the decay rate of the FWM signal at zero excitation density.
The zero-excitation-density decay rate of the FWM signal can be determined by extrapolating of excitation-density-dependence decay rate to zero-excitation-density \cite{moody2015intrinsic, boule2020coherent}.
We interpret the parameter $\gamma_0$ as the expected decay rate of the FWM signal in the hypothetical case where the inter-excitonic interactions are absent.

In this section we have provided a detailed comparison of the MOBE-based simulations and the AO model.
Our quantitative findings can be summarized as
\begin{subequations}
\begin{eqnarray}   
\dfrac{\xi}{\Delta}&=&{\dfrac{\gamma'}{\omega'}},\label{a}
\\
{\gamma_{01}(0)}& =&{\gamma_0}, \label{b}
\\
{\gamma_{EID}}&\propto& {\gamma'} \label{c}.
\end{eqnarray}
\end{subequations}
We have shown that both the models are consistent with each other.
Next, we compare them to the experimental data.

\begin{figure}[t]
\includegraphics{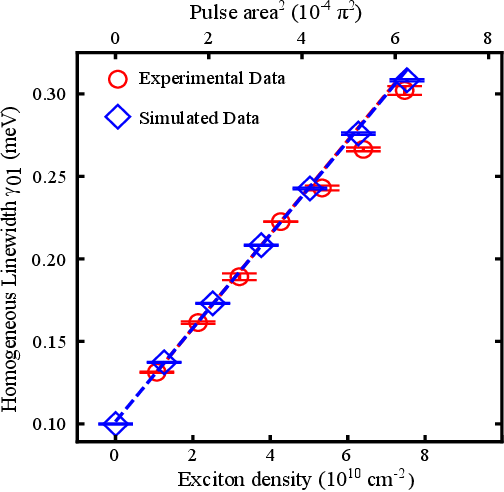}
\caption{Excitation-density dependence of the homogeneous linewidth $\gamma_{01}$ obtained by fitting the experimental data (red circles) and MOBE simulations (blue diamonds) to the AO model.
MOBE simulations were performed for $N\gamma'$ = 100 meV, $N\omega'$ = 150 meV.
The linear fits are shown by the blue and red dashed lines show an almost perfect overlap with each other.}
\label{fig:fig8_zerohomogeneous}
\end{figure}

\section{\label{sec:expt}Comparison with experiments}

\begin{figure}[t!]
\includegraphics{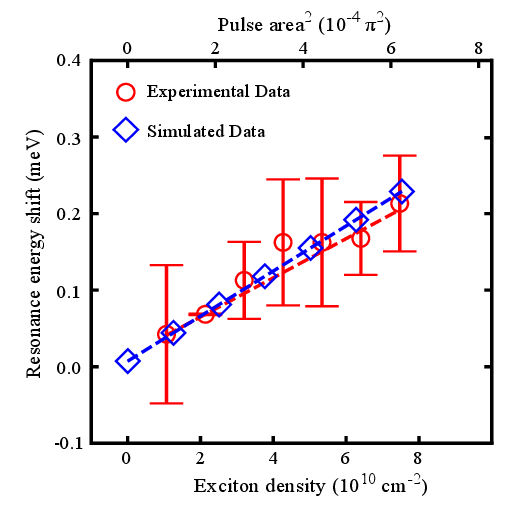}
\caption{Excitation-density dependence of the shift in the resonance frequency $\omega_{01}$ in experiments (red circles) and MOBE simulations (blue diamonds).
The data and simulations are the same as those used in Fig. \ref{fig:fig8_zerohomogeneous}.}
\label{fig:fig9_Peakshift}
\end{figure}

2DCS experiments were performed on GaAs QWs using the three-pulse noncollinear geometry \cite{bristow2009versatile}.
The sample comprising four periods of GaAs QWs with a width of 10 nm and Al$_{0.3}$Ga$_{0.7}$As barriers with the same width was kept at a temperature of 10 K.
These experiments were performed in the cocircular polarization scheme so that biexcitons are not excited \cite{bristow2009polarization}.
Excitation densities in the range of $\sim 1 - 8 \times 10^{10}$ cm$^{-2}$ per QW per excitation pulse were used.
The other experimental details were same as those used previously \cite{singh2016polarization}.

We fit the slices obtained from the experimental data with the AO model to obtain the parameters $\xi$, $\Delta$, $\gamma_{01}$, and $\omega_{01}$.
The observed dependence of $\gamma_{01}$ is shown in Fig. \ref{fig:fig8_zerohomogeneous} by red circles.
We repeated the experiment thrice at each power; the error bars for the fits to the experimental data indicate the standard deviation in the best-fit values.
We measure $\omega_{01}$ through the overall shift of the peak of the 2D spectra with changing excitation density; the shifts are shown in Fig. \ref{fig:fig9_Peakshift}.
These experimental findings were reproduced by solving the MOBEs for $N\gamma' = 100$ meV and $N\omega' = 150$ meV.
The simulated spectra are also fit with the AO model following the procedure discussed in Sec. \ref{sec:ao}.
The corresponding results are shown in Figs. \ref{fig:fig8_zerohomogeneous} and \ref{fig:fig9_Peakshift} with blue diamonds.
We have used an arbitrary scaling factor between the exciton density and the square of pulse area to show the correspondence between the fit data obtained from experiments and the MOBE simulations.
Similar fit parameters highlight that the lineshapes obtained from MOBEs are equivalent to the experimental data.
We note that in order to obtain a change in $\gamma_{01}$ that is equivalent to the experiment, we had to go beyond the $\chi^{(3)}$ regime.

Analogous to Eq. \ref{eq:4_gamma01}, we can quantify the excitation-density dependence of the resonance frequency through
\begin{equation}
\label{eq:omega01}
\omega_{01}(N_{x})= \omega_{01}{(0)} + \omega_{EIS} N_{x},
\end{equation}
where $\omega_{01}(0)$ is the resonance frequency of the $\ket{0}$$\leftrightarrow$$\ket{1}$ transition at zero excitation density and $\omega_{EIS}$ is the rate of change of the resonance frequency with the excitation density.
We did not discuss this phenomenon in Sec. \ref{sec:ao} because the small shifts in the resonance frequency were partially masked by the finite resolution of the simulations.
The increased pulse area and the $N\omega'$ term make the shift in resonance frequency clearly observable now.

Consistent with the earlier results, we did not see any excitation-density dependence on the symmetry-breaking parameters $\xi$ and $\Delta$ (data not shown) for either the experimental data or the simulation results.
Taking weighted averages, as discussed in Sec. \ref{sec:ao}, results in $\Delta = 19 \pm 3 \; \mu$eV and $\xi = 6 \pm 1 \; \mu$eV for the experiment with $\dfrac{\Delta}{\xi} = 3.2 \pm 0.7$.
The corresponding values for the MOBE simulations are $\Delta = 6.2 \pm 0.7 \; \mu$eV and $\xi = 4.3 \pm 0.4 \; \mu$eV corresponding to $\dfrac{\Delta}{\xi} = 1.4 \pm 0.2$; this should ideally be equal to 1.5 based on the ratio $\dfrac{\omega'}{\gamma'}$.
Although these values are quite different, $\dfrac{\Delta}{\xi} > 1$ suggests that  EIS is more dominant than EID in this sample.
Likely reasons for the quantitative anomaly could be the inability of the fitting procedure to perfectly separate the effect of EIS and EID and the large uncertainty $\sim 16\%$ in the experimentally obtained values.

\section{\label{sec:end}Conclusion}
We have presented a comprehensive and quantitative comparison of modeling MBIs using the MOBEs and the AO model.
These models introduce interactions to excitons when they are treated as fermions (for the MOBEs) and bosons (for the AO model).
Interestingly, these complimentary approaches result in identical lineshape of 2D spectra.
We have also established a direct correspondence between the MBIs parameters used in the MOBEs and the AO model.
These findings are further strengthened by reproducing experimental data for GaAs QWs using both the models independently.
This study provides a possible framework to compare the relative strength of MBIs among different excitonic systems such as QWs \cite{wachter2002coherent, webber2015role}, layered semiconductors \cite{boule2020coherent, katsch2020exciton, erkensten2021exciton} and perovskite nanomaterials \cite{wu2015excitonic, march2017four, thouin2019enhanced}.
It can also be relevant for the ongoing efforts to develop flexible semiconductor quantum dot lasers \cite{nielsen2004many, schneider2004excitation, monniello2013excitation} and TMDCs-based valleytronics devices \cite{sie2018valley, mahmood2018observation, cunningham2019resonant}, which are affected by MBIs.

\begin{acknowledgments}

\end{acknowledgments}

We thank Steven T. Cundiff for fruitful discussions.
We thank Takeshi Suzuki and S.T.C. for help in collecting the experimental data used in this work.
The authors acknowledge support from the Science and Engineering Research Board (SERB), New Delhi under Project No. SRG/2020/000822.
B.D. acknowledges the Ministry of Education, Government of India for support from the Prime Minister's Research Fellows (PMRF) Scheme.

P.K. and B.D. contributed equally to this work.

\bibliography{apssamp}

\end{document}